# Evidence-Based Comparison of Modularity Support Between Java and Object Teams

Arlindo Lima
CITI, Departamento de Informática
Faculdade de Ciências e Tecnologia
Universidade Nova de Lisboa
2829-516 Caparica, Portugal
+3512948536
anrl@fct.unl.pt

Miguel Goulão
CITI, Departamento de Informática
Faculdade de Ciências e Tecnologia
Universidade Nova de Lisboa
2829-516 Caparica, Portugal
+3512948536
miguel.goulao@di.fct.unl.pt

Miguel P. Monteiro
CITI, Departamento de Informática
Faculdade de Ciências e Tecnologia
Universidade Nova de Lisboa
2829-516 Caparica, Portugal
+3512948536
mmonteiro@di.fct.unl.pt

## ABSTRACT

**Background:** Aspect-oriented programming (AOP) is an emerging programming paradigm whose focus is about improving modularity, with an emphasis on the modularization of crosscutting concerns.

**Objective:** The goal of this paper is to assess the extent to which an AOP language – ObjectTeams/Java (OT/J) – improves the modularity of a software system. This improvement has been claimed but, to the best of our knowledge, this paper is the first attempting to present quantitative evidence of it.

**Method:** We compare functionally-equivalent implementations of the Gang-of-Four design patterns, developed in Java and OT/J, using software metrics.

**Results:** The results of our comparison support the modularity improvement claims made in the literature. For six of the seven metrics used, the OT/J versions of the patterns obtained significantly better results.

**Limitations:** This work uses a set of metrics originally defined for object-oriented (OO) systems. It may be the case that the metrics are biased, in that they were created in the context of OO programming (OOP), before the advent of AOP. We consider this comparison a stepping stone as, ultimately, we plan to assess the modularity improvements with paradigm independent metrics, which will conceivably eliminate the bias. Each individual example from the sample used in this paper is small. In future, we plan to replicate this experiment using larger systems, where the benefits of AOP may be more noticeable.

**Conclusion:** This work contributes with evidence to fill gaps in the body of quantitative results supporting alleged benefits to software modularity brought by AOP languages, namely OT/J.

## Categories and Subject Descriptors

D 2.2 [**Software Engineering**]: Design Tools and Techniques – *object-oriented design methods*.

D.2.8 [**Software Engineering**]: Metrics – *complexity measures, product metrics*.

D.3.3 [**Programming Languages**]: Language Constructs and Features – *classes and objects, patterns and polymorphism.*

## 1. MOTIVATION
### 1.1 Problem Statement

AOP is an emerging software composition paradigm whose main purpose is to improve modularity in software, when compared to traditional programming paradigms like OOP, with a strong emphasis on the modularization of crosscutting concerns [22]. AspectJ [21, 35] is the best-known AOP representative and seems to be the most widely used. However, many consider AspectJ to have a negative impact on software modularity [6, 14, 28, 30].

OT/J [16, 18, 32] is a more recent AOP language and in [16] Herrmann claims that concepts and mechanisms from OT/J *"provide a better decoupling, modularization and flexibility"* than AspectJ. However, alleged superiorities are mostly supported by argumentation. So far, systematic studies and quantitative evidence supporting such claims are lacking.

This paper presents an initial exploratory study of OT/J's impact on programs' modularity, focused on comparing results obtained by Java and OT/J. Comparisons with AspectJ are ongoing and are left for future work. The study was carried out through use of the metrics suite offered by the Eclipse [36] plug-in for developing OT/J (OTDT) [33] that collects metrics for both Java [31] and OT/J.

This paper's organization is adapted from the "standard" experimental report structure proposed in [20]. This section states the problem of quantitatively assessing OT/J's support for modularity. Section 2 discusses relevant related work performed on the quantitative assessment of other AOP languages with respect to their support for modularity. Section 3 presents a short overview of OT/J. Section 4 discusses the design of our empirical evaluation of OT/J's support for modularity, in contrast with that of Java. Section 5 presents the execution of the empirical study. Section 6 reports the results of that study. Section 7 discusses the results. Section 8 concludes the paper and outlines our plans for future work.

## 1.2 Research Objectives
Research objectives are presented in the format proposed in [20]:

**Analyze** the OT/J language,
**for the purpose of** assessing the usefulness of its language constructs (using the Java language as a yardstick),
**with respect to** software modularity,
**from the point of view of** developers who may implement analogous systems in both the ObjectTeams/Java and Java languages,
**in the context of** an introductory observational study on a repository that includes functionally equivalent pattern implementations in both Java and OT/J.

## 1.3 Context
This study builds on previous work within our research group [11, 12], namely the implementation of the well-known Gang-of-Four (GoF) design patterns [9] in OT/J[1]. Two repositories of implementations of the GoF are used: Hannemann & Kiczales' [15] and Cooper's [7]. We consider results from this study valid only in the context of the patterns' examples used, rather than as applying to software modules in general. Further research must be conducted to assess which conclusions are specific to the implementations used and which are generalisable.

## 2. RELATED WORK
Few metrics for AOP were proposed in the literature. Zhao was one of the first to propose modularity metrics specific for AOP, having formalised coupling [37] and cohesion [38] metrics. Sant'Anna *et al.* developed a metrics' suite which includes metrics adapted from known OOP metrics [27]. Several quantitative studies using this suite have been developed. Garcia *et al.* [10] performed a quantitative assessment of the modularity of design-patterns examples in AspectJ, comparing them to Java implementations of the same examples. Kulesza *et al.* [23] studied the effect of AspectJ with respect to maintainability. Both studies were favourable to AOP.

Lopes and Bajracharya [24] used *Design Structure Matrices* and *Net Option Value* to compare AOP and OOP systems. Their work suggests that, in some cases, AOP is beneficial, while in others it should be considered prejudicial. Similar mixed results were obtained by Bryton [3].

Performing comparisons between two paradigms is problematic: it's easy to mix apples with oranges, particularly because each paradigm uses its own language mechanisms to support the features under scrutiny (e.g. modularity). A possible solution is to develop *paradigm-independent metrics*. An example is provided by Bryton and Brito e Abreu [4], where a paradigm independent meta-model for modularity is proposed and a set of metrics is formally defined upon the paradigm-independent meta-model. A related approach is to develop a multiparadigm metric, that is, a metric that measures concepts from multiple paradigms [25]. In our opinion, this approach is more prone to introducing biases than the paradigm-independent one. It mixes concepts from each of the paradigms in the same metric, rather than translating those concepts to an allegedly neutral representation before measuring them. In both approaches, the challenge is to ensure that the mapping from each paradigm to the paradigm-independent (or the multi-paradigm) representation is "fair". In this context, "fairness" means that mappings between different paradigms do not artificially introduce any sort of bias in the metrics values. Otherwise, significant differences observed in the metrics may result from the mapping, rather than from fundamental differences introduced by each of the paradigms, as desirable in a metrics-based paradigm comparison.

## 3. AN OVERVIEW OF OBJECT TEAMS
This section outlines the main features of OT/J [18], which is the implementation of the Object Teams model for Java. Refer to [18] for an exhaustive definition of the language.

Object Teams introduces a new module concept, the *Team*, which unifies the notions of class and package (and can be seen as an aspect module). A Team can contain one or more *Roles*. A Role encapsulates behaviour which can decorate one base class (in this case, the Role is considered to be bound to the base class). Teams and Roles can be seen, respectively, as outer and inner Java classes.

The code sketch in Listing 1 illustrates these concepts. `PrinterAdapterTeam` is a Team and `Adapter` is a Role bound to the base class `SOPrinter`.

```
01  public team class PrinterAdapterTeam {
02      public class Adapter playedBy SOPrinter {
03          // Role implementation
04      }
05      // remaining Team implementation
06  }
07  public class SOPrinter {
08      // normal (base) class
09  }
```
**Listing 1. Examples of a Team, a Role and a base class.**

The binding of a Role class to a base class has no effect on its own, but is the basis for 2 kinds of bindings: *callins* and *callouts*.

### 3.1 *Callin* Binding
A *callin* binding declares that a given Role method should be executed for every call of the associated base method (line 4 of Listing 2). This type of binding can be of type *before*, *after* or *replace*. This is similar to *advice* in AspectJ but it is worth noting that this mechanism retains a polymorphic feel, with each individual *callin* mapping being one-to-one. A parallel can also be made in the context of traditional inheritance between the way a subclass constructor implicitly calls a superclass constructor.

In Listing 2, after the execution of `displayMsg`, the *callin* method `updateObservers` will be invoked:

```
01  public team class ScreenObserverTeam extends
02          ObserverProtocolTeam {
03      public class Subject playedBy Screen {
04          updateObservers <- after displayMsg;
05          // void updateObservers() inherited from the Subject
06          // Role of the super-Team ObserverProtocolTeam
07      }
08  }
09  public class Screen {
10      public void displayMsg(String s) {
11          print(s);
12      }
13  }
```
**Listing 2. Example of a *callin* binding**

---
[1] The material used for this study is available at:
http://ctp.di.fct.unl.pt/~mpm/AOLA/

## 3.2 *Callout* Binding

*Callout* bindings allow Role instances to forward method calls to base methods (or fields). This can be used to "implement" abstract methods of a Role (see lines 5 and 6 in Listing 3 for an example) in a way that mimics the relationship between abstract classes and concrete subclasses in traditional inheritance. This way, a Role can contain abstract methods and still be concrete, completed through *callouts* to the base. This mechanism is not present in AspectJ.

```
01  public team class ScreenObserver {
02      public class Observer playedBy Screen {
03          public abstract void update();
04          public abstract int howMany();
05          update -> refresh;
06          howMany -> get elems;
07      }
08  }
09  public class Screen {
10      public int elems;
11      public void refresh() {
12          //implementation
13      }
14  }
```

**Listing 3. Example of a *callout* binding.**

## 3.3 Translation polymorphism

There is no sub-type relation between a Role and its base class but under certain conditions their instances are substitutable. Two mechanisms allow this kind of polymorphism: *lifting* (translation of a base class to one of its Roles) and *lowering* (the inverse of *lifting*, i.e., the mapping of a Role to its associated base).

## 3.4 Team Inheritance

In OT/J, Teams and Roles are first class citizens, so inheritance works as traditionally for both Teams and Roles, with respect to their members. Roles enclosed within a super-Team are inherited by sub-Teams via *implicit inheritance*. Thus, if a sub-Team has a Role of the same name as an inherited Role, the latter is implicitly overridden and subject to dynamic dispatch.

## 3.5 Other features

OT/J offers several other features like the possibility to dynamically activate/deactivate Teams (which determines the effectiveness of a *callin* bindings) and *decapsulation*, i.e., the violation of access restrictions to bind Roles to otherwise inaccessible (e.g., private) base methods and fields. Herrmann argues that this mechanism is useful for extending an existing module with new functionalities, without having to modify it – but warns that it should be used as a last resort [17].

## 4. EXPERIMENTAL DESIGN
### 4.1 Goals

The research objective presented in sub-section 1.2 is too abstract for the purposes of the proposed assessment. To make it more concrete, we break it down into seven sub-goals (Table 1), where the variation lies on the metric under assessment. In the sub-goals definition "(...)" is used to denote that we keep the corresponding part of the more abstract research objective. This allows us to highlight the differences among the seven sub-goals.

**Table 1. Research goals.**

| Goal | Description |
|---|---|
| G1 | Analyze the OT/J language, (...) with respect to the *Coupling between object classes* [5] metric, (...) |
| G2 | Analyze the OT/J language, (...) with respect to the *Number of classes used by this class* metric, (...) |
| G3 | Analyze the OT/J language, (...) with respect to the *Number of classes using this class* metric, (...) |
| G4 | Analyze the OT/J language, (...) with respect to the *Response For a Class* [5] metric, (...) |
| G5 | Analyze the OT/J language, (...) with respect to the *Number of Children* [5] metric, (...) |
| G6 | Analyze the OT/J language, (...) with respect to the *Depth of Inheritance Tree* [5] metric, (...) |
| G7 | Analyze the OT/J language, (...) with respect to the *Lack of Cohesion in Methods* [5] metric, (...) |

### 4.2 Experimental Units

The experimental units of this observational study are the individual examples of the aforementioned design patterns implementations.

### 4.3 Experimental Material

- All the 23 Hannemann & Kiczales' [15] GoF design patterns [9] examples, implemented in both OT/J and Java.
- 18 of James Cooper's [7] GoF design patterns examples, implemented in both OT/J and Java.

Five of the James Cooper's patterns (*Builder*, *Façade*, *Factory Method*, *Interpreter* and *State*) were not yet implemented in OT/J. We're planning to implement them in the future.

### 4.4 Tasks

As noted on the previous sub-section, the subjects of this study are design pattern implementations. As such, this common item in the experimental design description is not applicable for this study.

## 4.5 Hypotheses and variables

### 4.5.1 Hypotheses
The goals lead us to test seven different basic hypotheses, in order to assess the effect of OT/J on each metric (when compared to Java). We identify the hypotheses as *H1*, *H2*, *H3*, *H4*, *H5*, *H6* and *H7* (Table 2). For each hypothesis, we formulate both a null and an alternative hypothesis.

**Table 2. Research hypotheses.**

| Hypotheses | | |
|---|---|---|
| H1 | $H1_0$ | OT/J provides no significant improvement on the patterns' *Coupling between object classes*. |
| | $H1_1$ | OT/J provides a significant improvement on the patterns' *Coupling between object classes*. |
| H2 | $H2_0$ | OT/J provides no significant improvement on the patterns' *Number of classes used by this class*. |
| | $H2_1$ | OT/J provides a significant improvement on the patterns' *Number of classes used by this class*. |
| H3 | $H3_0$ | OT/J provides no significant improvement on the patterns' *Number of classes using this class*. |
| | $H3_1$ | OT/J provides a significant improvement on the patterns' *Number of classes using this class*. |
| H4 | $H4_0$ | OT/J provides no significant improvement on the patterns' *Response For a Class*. |
| | $H4_1$ | OT/J provides a significant improvement on the patterns' *Response For a Class*. |
| H5 | $H5_0$ | OT/J provides no significant improvement on the patterns' *Number of Children*. |
| | $H5_1$ | OT/J provides a significant improvement on the patterns' *Number of Children*. |
| H6 | $H6_0$ | OT/J provides no significant improvement on the patterns' *Depth of Inheritance Tree*. |
| | $H6_1$ | OT/J provides a significant improvement on the patterns' *Depth of Inheritance Tree*. |
| H7 | $H7_0$ | OT/J provides no significant improvement on the patterns' *Lack of Cohesion in Methods*. |
| | $H7_1$ | OT/J provides a significant improvement on the patterns' *Lack of Cohesion in Methods*. |

### 4.5.2 Independent variables
The independent variable is the same for all the hypotheses. This variable, which we'll call "**Is OT/J**" assumes the value `true` for pattern instances implemented in OT/J and `false` otherwise.

### 4.5.3 Dependent variables
The dependent variables used in this experiment represent the various metrics collected with the OTDT plug-in. Only metrics that could be applied to both Java and OT/J were used. However, two sets of metrics were intentionally left out, despite being computed for both languages: Lines of Code (due to being known poor predictors of modularity) and Number of Classes/Interfaces (due to OT/J Teams and Roles not being counted for these metrics and hence rendering them not directly comparable). Except for the *Depth of Inheritance Tree* metric, all our dependent variables are normalized on the experimental unit's number of modules (Classes, Teams, Roles and Implementations) for which the dependent variables apply. This is to mitigate the effect of the implementations' different sizes.

In summary, our dependent variables are:

*Coupling between object classes (CBO)*. The OTDT defines this metric as the number of classes coupled to a class X through a *uses* or *used by* relationship (Teams, Roles and Interfaces are also included in this definition of *class*), and, according to source code comments found in [34], is implemented as defined in [5]. OTDT offers two variations of this metric, of which "*closed scope*" was selected to force the computation of CBO exclusively for the classes composing each experimental unit.

*Number of classes used by this class (NCUBC)*. Defined, by OTDT, as the number of classes *used by* a class X (Teams, Roles and Interfaces are also included in this definition of *class*). The "*closed scope*" variation is used.

*Number of classes using this class (NCUC)*. Defined, by OTDT, as the number of classes *using* a class X (Teams, Roles and Interfaces are also included in this definition of *class*). The "*closed scope*" variation is used.

*Response For a Class (RFC)*. The OTDT defines this metric as the cardinality of the set containing the methods declared by and the methods called by a class X (Teams and Roles are also included in this definition of *class*), and, according to source code comments found in [34], is implemented as defined in [5]. The "*closed scope*" variation is used.

*Number of Children (NOC)*. Defined, by OTDT, as the number of immediate subtypes of a type X and, according to source code comments found in [34], is implemented as defined in [5]. The "*closed scope, include implementation*" variation is used (to count the implementations of an Interface as its children).

*Depth of Inheritance Tree (DIT)*. Defined, by OTDT, as the number of super-types in the longest path from a type X to a root type of its inheritance hierarchy and, according to source code comments found in [34], is implemented as defined in [5]. The "*closed scope, include implementation*" variation is used.

*Lack of Cohesion in Methods (LCOM)*. The OTDT's definition for this metric is: number of method pairs that do not have any used instance variables in common, minus the number of method pairs that have at least one used instance variable in common. The minimum value for this metric is zero. According to source code comments found in [34], LCOM is implemented as defined in [5]. As RFC, this metric doesn't apply to Interfaces.

## 4.6 Design
In this case study, we use a single group of subjects (the pattern implementations) and a single observation.

## 4.7 Procedure
Collection of metrics was automated, using the built-in metrics available in the OTDT Eclipse plug-in (version 1.4.0 Milestone 2, based on Eclipse version 3.6.0 M4).

## 4.8 Analysis Procedure
We follow the following steps:

- **Compute descriptive statistics**: For all our independent and dependent variables, we collect a set of descriptive statistics. These descriptive statistics provide us with a first overview of our data, which we further detail in subsequent analyses.

- **Normality tests**: Data is checked for normality, so that the statistics tests which are suitable for our data can be selected.
- **Analysis of differences between groups**: Finally, we perform a test to detect whether there are significant differences between groups. This allows us to test the hypotheses stated in sub-section 4.5.1.

## 5. EXECUTION
### 5.1 Sample
We used the metrics set on the experimental material. No exclusions were made.

### 5.2 Preparation
No special preparations were required, other than installing a version of the OTDT shipped with the metrics plug-in. The implementations used in [11, 12] were analyzed as they were, with no adaptations specifically for the present study.

### 5.3 Data Collection Performed
Data collection follows the plan outlined in the sub-section 4.7.

## 6. ANALYSIS
### 6.1 Descriptive Statistics
For each variable, we present the number of cases, the mean value within the sample, standard deviation, minimum value, maximum value, skewness and kurtosis (Table 3).

**Table 3. Descriptive statistics of the metrics.**

|        | *H1*   | *H2*   | *H3*   | *H4*   | *H5*    | *H6*  | *H7*   |
|--------|--------|--------|--------|--------|---------|-------|--------|
| Metric | CBO    | NCUBC  | NCUC   | RFC    | NOC     | DIT   | LCOM   |
| N      | 82     | 82     | 82     | 82     | 82      | 82    | 82     |
| Mean   | 1,6395 | 0,8524 | 0,8524 | 3,8468 | 0,2319  | 0,79  | 0,1327 |
| Std.Dv.| 0,5178 | 0,2802 | 0,2802 | 1,5932 | 0,2064  | 0,623 | 0,5712 |
| Min.   | 0,67   | 0,33   | 0,33   | 1,25   | 0,00    | 0     | 0,00   |
| Max.   | 3,43   | 1,71   | 1,71   | 8,67   | 0,67    | 3     | 5,00   |
| Skew.  | 0,916  | 0,838  | 0,838  | 0,731  | 0,362   | 0,486 | 7,880  |
| Kurt.  | 1,357  | 0,870  | 0,870  | 0,280  | -1,153  | 0,526 | 67,057 |

To decide whether it is appropriate to use parametric tests for our hypothesis, we need to check if the variable has a normal distribution. Positive skewness indicates an asymmetric distribution, with a higher frequency of the variable's lower values. In other words, the distribution is right-skewed. This contrasts with the normal distribution, which is symmetric and should therefore exhibit a skewness of 0, providing us a hint on the non-normality of our data.

We use further tests to confirm the non-normality of this variable. Table 4 presents results of two such tests: the Kolmogorov-Smirnov with the Lilliefors correction and the Shapiro-Wilk's normality tests. The former is the most widely used test and adequate for our sample size. The latter is often used with smaller samples, and used here for confirmation purposes only. The null hypothesis, for each of the tests, is that the sample comes from a normal distribution. The alternative is that the sample comes from a non-normal distribution.

**Table 4. Normality tests.**

|        | Kolmogorov-Smirnov | | | Shapiro-Wilk | | |
|--------|---------|----|-------|---------|----|-------|
| Metric | Statist | df | Sig.  | Statist | df | Sig   |
| CBO    | 0,152   | 82 | 0,000 | 0,945   | 82 | 0,002 |
| NCUBC  | 0,147   | 82 | 0,000 | 0,947   | 82 | 0,002 |
| NCUC   | 0,147   | 82 | 0,000 | 0,947   | 82 | 0,002 |
| RFC    | 0,106   | 82 | 0,024 | 0,960   | 82 | 0,011 |
| NOC    | 0,187   | 82 | 0,000 | 0,893   | 82 | 0,000 |
| DIT    | 0,325   | 82 | 0,000 | 0,756   | 82 | 0,000 |
| LCOM   | 0,408   | 82 | 0,000 | 0,226   | 82 | 0,000 |

These values confirm the non-normality of the sample. As such, we use non-parametric procedures for testing our hypotheses.

### 6.2 Data Set Reduction
No experimental units were removed from the sample.

### 6.3 Hypotheses testing
We perform the Mann-Whitney U test (Table 6), which is a non-parametric alternative to assess whether two samples of observations come from the same population. The test starts by ranking all observations, regardless of the sample they come from. Values are ranked in descending order. Table 5 summarizes the information concerning computed ranks. Note that the number of implementations per language is constant (41) and that all OT/J's mean ranks are lower.

**Table 5. Ranks for the hypotheses.**

| H   | Metric | Lang. | N  | Mean Rank | Sum Ranks |
|-----|--------|-------|----|-----------|-----------|
| *H1* | CBO   | Java  | 41 | 48,44     | 1986,00   |
|     |        | OT/J  | 41 | 34,56     | 1417,00   |
| *H2* | NCUBC | Java  | 41 | 47,20     | 1935,00   |
|     |        | OT/J  | 41 | 35,80     | 1468,00   |
| *H3* | NCUC  | Java  | 41 | 47,20     | 1935,00   |
|     |        | OT/J  | 41 | 35,80     | 1468,00   |
| *H4* | RFC   | Java  | 41 | 51,35     | 2105,50   |
|     |        | OT/J  | 41 | 31,65     | 1297,50   |
| *H5* | NOC   | Java  | 41 | 56,33     | 2309,50   |
|     |        | OT/J  | 41 | 26,67     | 1093,50   |
| *H6* | DIT   | Java  | 41 | 47,83     | 1961,00   |
|     |        | OT/J  | 41 | 35,17     | 1442,00   |
| *H7* | LCOM  | Java  | 41 | 43,32     | 1776,00   |
|     |        | OT/J  | 41 | 39,68     | 1627,00   |

The Mann-Whitney U (M-W U) tests are summarized in Table 6. We can observe that for hypotheses *H2* and *H3*, the null hypothesis can be rejected with $p < 0,05$. Hypotheses *H1*, *H4*, *H5* and *H6* can also be rejected with $p < 0,01$. In other words, for these six hypotheses we found significantly lower metric values for the OT/J instances, comparing to their Java counterparts.

For hypothesis *H7*, no significant differences were found, so we can reject it.

Table 6. Mann-Whitney U test.

| H | Metric | M-W U | Wil. W | Z | AS(2-t) |
|---|--------|-------|--------|---|---------|
| *H1* | CBO | 556,000 | 1417,000 | -2,645 | 0,008 |
| *H2* | NCUBC | 607,000 | 1468,000 | -2,169 | 0,030 |
| *H3* | NCUC | 607,000 | 1468,000 | -2,169 | 0,030 |
| *H4* | RFC | 436,500 | 1297,500 | -3,747 | 0,000 |
| *H5* | NOC | 232,500 | 1093,500 | -5,735 | 0,000 |
| *H6* | DIT | 581,000 | 1442,000 | -2,789 | 0,005 |
| *H7* | LCOM | 766,000 | 1627,000 | -0,901 | 0,368 |

## 7. INTERPRETATION
### 7.1 Evaluation of Results and Implications
Our results support the claims on the improved modularity brought by OT/J [16]. Six out of the seven metrics used to assess the modularity (CBO, NCUBC, NCUC, RFC, NOC and DIT) showed significantly better values for the OT/J implementations.

*Coupling between object classes* (CBO) measures the number of classes coupled to each class of the system. Since low coupling is a desirable feature of a modular design, this could indicate a strength of OT/J. A possible explanation could reside on two mechanisms introduced by OT/J, which CBO ignores: *callins* and *callouts*. Roles and base classes coupled via these bindings are ignored by this metric. This makes OT/J implementations' measured coupling potentially lower than it is in reality. It must be noted that this metric is criticized for its lack of differentiation between coupling within a component and inter-component [2, 19].

The *Number of classes used by this class* (NCUBC) *and Number of classes using this class* (NCUC) metrics are related to the coupling of a system (in fact, CBO uses these two metrics). As aforementioned, low coupling is a desirable feature of a modular system, which could, once again, indicate a strong point for OT/J. However, the considerations made for CBO apply likewise to these two metrics, due to *callins* and *callouts* not being counted for NCUBC and NCUC. The values of these two metrics are always equal because a "closed scope" version is used (note that if A *uses* B, then B is *used by* A).

*Response For a Class* (RFC), defined as the cardinality of the set containing the methods declared by and the methods called by a class, is improved by OT/J, as well. Originally defined as a complexity metric [5], RFC is highly correlated to the coupling of a system [29], being therefore also related to its modularity. *Callins* and *callouts* are not counted for this metric, which, in many implementations, originates Roles with a value of zero RFC (Roles with bindings, but no methods – all its methods are inherited). With the normalization of this metric, these zero RFC Roles have the negative effect of reducing the value of this metric for the whole OT/J experimental unit.

*Number of Children* (NOC) counts how many direct subtypes a given entity has. Based on NOC, it is clear that OT/J significantly reduces the use of inheritance as an extension mechanism. This can be justified by the usage of a new extension mechanism introduced by OT/J: the *played by* relationship, through which a base class can see its behaviour extended by a Role and vice-versa. This mechanism is ignored by the metric.

*Depth of Inheritance Tree* (DIT) measures how many super-types there are in the longest path from a type X to a root type of its inheritance hierarchy. In the context of OOP, a high DIT was shown to be correlated to a complex design and fault-prone code [8]. However, the lower value obtained in OT/J implementations may be a reflection of the limitation discussed above: OT/J's richer set of mechanisms for module extensibility are not fully assessed.

With respect to *Lack of Cohesion in Methods* (LCOM), no significant difference was found. As LCOM has no discriminative power in this context, no conclusions concerning differences between the two languages can be drawn from it.

### 7.2 Threats to Validity
We consider two kinds of threats: the first relates to a potential bias introduced by the metrics suite used in this study and the second to the experimental units used.

The metrics suite was originally developed for OOP and later adapted to AOP, and in particular to OT/J. As discussed in [3, 4], performing comparisons among different paradigms with metrics defined especially for one of those paradigms may yield misleading results. Although these metrics have been validated in the context of OOP, their applicability to AOP is yet to be demonstrated. In our opinion, we would ideally strive for a paradigm-independent set of metrics when performing these kinds of inter-paradigm assessments. However, paradigm-independent metrics are still in an "embryonic" stage.

As for the external validity of our experiment, we should note that the design pattern implementations, in both languages, are fairly small. This is especially relevant, as modularity is a quality attribute that becomes increasingly important as systems grow larger and more complex. In addition, our implementation set is composed solely of design pattern implementations and, in that sense, may lack heterogeneity. This may potentially introduce biases related to idiosyncrasies of this group. Furthermore, the implementations were built by a small number of developers and may therefore be tainted with their personal style, although special effort was made, as much as possible, to mitigate this effect.

### 7.3 Inferences
The analysis performed in this observational study should hold for implementations of similar characteristics (in particular their complexity). Extrapolating these results to larger implementations should be performed with caution, as discussed in the previous section.

### 7.4 Lessons Learned
With respect to the operationalisation of the observational study itself, there were a few challenges to overcome, to ensure the data quality for our analysis.

The interoperability of the metrics collection tool was insufficient, as no export feature for the metrics results was available. This resulted in a time-consuming and error-prone manual copying of the results to a statistical analysis tool, which would make this experience unfeasible for larger data sets. This problem can be overcome with the inclusion of such a feature in a future version of the tool.

To assure a fair comparison of the different sets of pattern implementations, some refactoring was carried out prior to the study, due to unsuitable coding style in one repository. This allowed us to factor out potential discrepancies introduced by different levels of expertise of the developers of the pattern examples. While this heterogeneity would be a desirable feature in a larger study (thousands of implementations), not performing those refactorings would introduce noise in our sample, due to the relatively small number of developers involved in it and to the fact that each developer typically contributed either to the Java or the OT/J repository, but not to both.

## 8. CONCLUSIONS AND FUTURE WORK

### 8.1 Summary

The results of this experiment provide evidence supporting the claims of improved modularity in OT/J implementations. However, we do not regard this as definitive evidence in favour of those claims. 6 out of the 7 metrics employed in this study were favourable to the OT/J implementations and one was inconclusive. However, this might be explained by the metrics obliviousness to some of the new OT/J's mechanisms. Another possible shortcoming of these metrics is that they have not been proved to be paradigm independent in the past, which may potentially introduce bias. One of the main contributions of this work is that this is, to the best of our knowledge, the first quantitative study to assess OT/J with respect to its support to modularity. A second contribution concerns the critical evaluation of the use of OOP metrics, when applied to another paradigm. The impact of that paradigm shift, on the set of metrics, is yet to be fully understood.

### 8.2 Impact

With AOP technology gaining importance in current software system developments, as well as on the evolution of legacy systems, the community should be made aware of the lack of quantitative evidence supporting the alleged benefits of AOP. Our results are consistent with these type of claims, but we believe more efforts in the experimental assessment of those claims are necessary. In particular, it may be the case that some claims only hold for particular domains. Experimentally identifying those domains would help practitioners to make informed and sound decisions concerning the effective usage of AOP.

### 8.3 Future Work

Most of the existing quantitative studies involving AOP focus on comparing AOP with OOP, usually using AspectJ and Java as paradigm representative languages. To the best of our knowledge, no quantitative studies explicitly comparing two instances of AOP languages have been published so far, although there have been studies using more than one AOP language – for instance, in the study reported in [13] AspectJ and CaesarJ [1, 26] were used. However, it was made clear that the study was geared to comparisons across paradigms, not between different AOP languages. In future, we plan to contribute in filling this gap by developing studies similar to the one described in this paper so as to cover multiple AOP languages.

Ongoing work includes a rigorous comparison between different instances of the AOP paradigm (e.g. AspectJ, OT/J and CaesarJ). This way, potentially important insights may be derived about the relative advantages and disadvantages of the various AOP languages, thus contributing to mature the AOP paradigm. In addition we plan to work on the development of paradigm independent metrics.

## 9. ACKNOWLEDGMENTS
The authors would like to thank CITI for the support received during this research.